\documentclass[12pt]{iopart}
\usepackage{graphicx}
\usepackage{setstack}
\usepackage{iopams}
\usepackage[tight]{subfigure}
\newcommand{\half}{\frac{1}{2}}

\newcommand{\dm}{Dzyaloshinskii-Moriya}
\newcommand{\ve}[1]{\mathbf{#1}}

\newcommand{\ket}[1]{\left| #1\right>}

\newcommand{\Eq}[1]{(\ref{#1})}
\newcommand{\Fig}[1]{figure~\ref{#1}}
\newcommand{\Tab}[1]{table~\ref{#1}}

\newcommand{\lch}{\ket{\swarrow\searrow}}
\newcommand{\rch}{\ket{\searrow\swarrow}}

\begin{document}

\title{\dm\ interaction in transport through single molecule transistors 
}
\author{S. Herzog$^{(1,2,3)}$, M. R. Wegewijs$^{(1,2,3)}$}
\address{$^1$ Institut f\"ur Theoretische Physik A,
      RWTH Aachen, 52056 Aachen,  Germany}
 \address{$^2$ Institut f\"ur Festk{\"o}rper-Forschung - Theorie 3,
      Forschungszentrum J{\"u}lich, 52425 J{\"u}lich,  Germany }
  \address{$^3$ JARA - Fundamentals of Future Information Technology}
\ead{m.r.wegewijs@fz-juelich.de}
\maketitle
\begin{abstract}

  The \dm\ interaction is shown to result in a canting of spins
 in a single molecule transistor.
  We predict non-linear transport signatures of this effect induced by spin-orbit coupling
  for the generic case of a molecular dimer.
  The conductance is calculated using a master equation and is found to exhibit a non-trivial dependence on the magnitude and direction of an external magnetic field.
  We show how three-terminal transport measurements allow for a determination of the coupling-vector characterizing the \dm \ interaction.
In particular, we show how its orientation, defining the intramolecular spin chirality, can be probed with ferromagnetic electrodes.
\end{abstract}
\submitto{\NT}
\section{Introduction}
Antisymmetric exchange interaction first identified by Dzyaloshinskii based on symmetry arguments~\cite{Dzyaloshinsky58a}, and derived by Moriya~\cite{Moriya60a}, is fundamental to the understanding of  weak ferromagnetism in materials like  $\alpha-\mathrm{Fe}_2\mathrm{O}_3$~\cite{Moriya63a}.
Nowadays such spin-orbit effects can be addressed in nanoscale experiments where transport plays an important role,
for instance, for STM on atomic chains or even single magnetic atoms on a surface~\cite{Rudenko09a}.
In particular, the \dm \ (DM) interaction favours spiralling spin-structures in systems with broken inversion symmetry, which is generically the case on a surface~\cite{Bode07a}.
Intrinsic effects of spin-orbit coupling in magnetic molecules have also been investigated in three terminal devices~\cite{Heersche06,Jo06,Grose08},
focusing on easy-axis magnetic anisotropy and quantum-spin tunnelling effects~\cite{Romeike06b,Lehmann07a}.
Whereas the latter are induced in second order in the spin-orbit (SO) interaction,
the DM interaction already arises in linear order.
In particular, DM interaction is known to have a large effect on the magnetism of dimer molecules~\cite{Miyahara08a} and plays an important role in the single molecular magnets $\mathrm{Mn}_6$ \cite{Bahr08a}, $\mathrm{Mn}_{12}$ \cite{Katnelson99a,DeRaedt04a,Wernsdorfer08a} and $\mathrm{V}_{15}$ \cite{DeRaedt04a,Konstantinidis02a}.
\par
For the electronic spectroscopy of single molecule transistors (SMTs)~\cite{vanderZant06} the spin-orbit interaction is of special importance, since it may generate a specific electric or magnetic field dependence of the transport~\cite{Gonzalez07}. This facilitates the experimental discrimination between the molecule's contribution to the electric current and that of possible spurious particles in a junction.
It also leads to a violation of spin-selection rules, which clearly affects the non-equilibrium occupations of the magnetic molecular states~\cite{Romeike06b}.
Since the \dm\ interaction relies fundamentally on broken spatial inversion symmetry,
which is typically present in a molecular junction,
it is of particular interest to identify its characteristic transport signatures.
\par
\begin{figure}[t!]
\centering
  \includegraphics[height=0.35\linewidth]{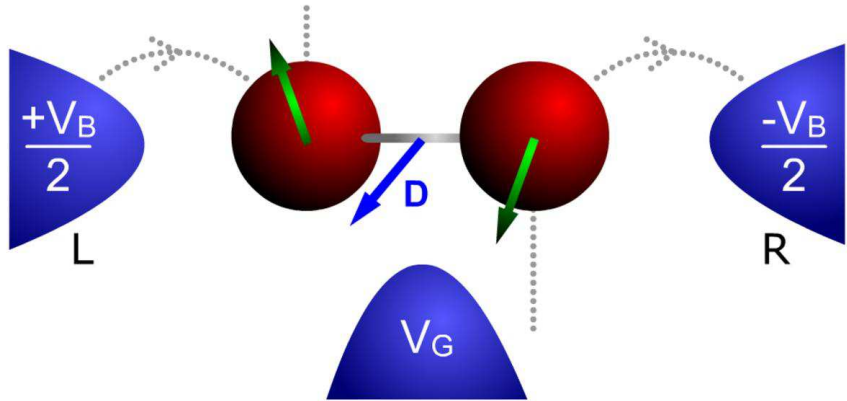}
  \caption{
    Dimer with localized spins, which are slightly canted due to interplay between DM interaction and isotropic exchange coupling $J$.
    The dimer is probed in a three terminal junction ($V_B$: Bias voltage, $V_G$: Gate voltage). Tunnelling is only possible between monomers and from one monomer to the adjacent lead. The grey vertical lines indicate the spin orientation without DM interaction.
  }\label{fig:orbconf}
\end{figure}

\dm\ interaction arises in its most elementary form in a molecular dimer~\cite{Miyahara08a} where two localized electrons interact via Coulomb and hybridization terms, in combination with local spin-orbit scattering into excited orbitals.
As a result, an effective spin-spin interaction arises between the spin 1/2 localized at the monomers labelled by $i=1,2$ (see \Fig{fig:orbconf}):
\begin{eqnarray}
  H_{\mathrm{eff}} & = &
   J \mathbf{S}_1 \cdot \mathbf{S_{}}_2
  + \mathbf{D} \cdot \left( \mathbf{S_{}}_1 \times \mathbf{S}_2 \right)
  + \mathbf{B} \cdot \left( \mathbf{S}_1 + \mathbf{S}_2 \right)
  \label{eq:eff}
\end{eqnarray}
The isotropic exchange typically dominates, resulting in a singlet and triplet split by $J$.
In three terminal molecular devices~\cite{Osorio07b,Roch08a} such a splitting in the meV range has been reported.
The antisymmetric exchange interaction is specified by the DM vector $\mathbf{D}$ which is an intrinsic property of the molecule. This interaction typically presents a weak perturbation which mixes the two multiplets of different total spin length, $S=0$ and $1$ here, resulting in small splittings $ \sim |\mathbf{D}|^2 /J \ll J$.
\begin{figure}[b!]
\centering
  \subfigure[]{
    \includegraphics[height=0.31\linewidth]{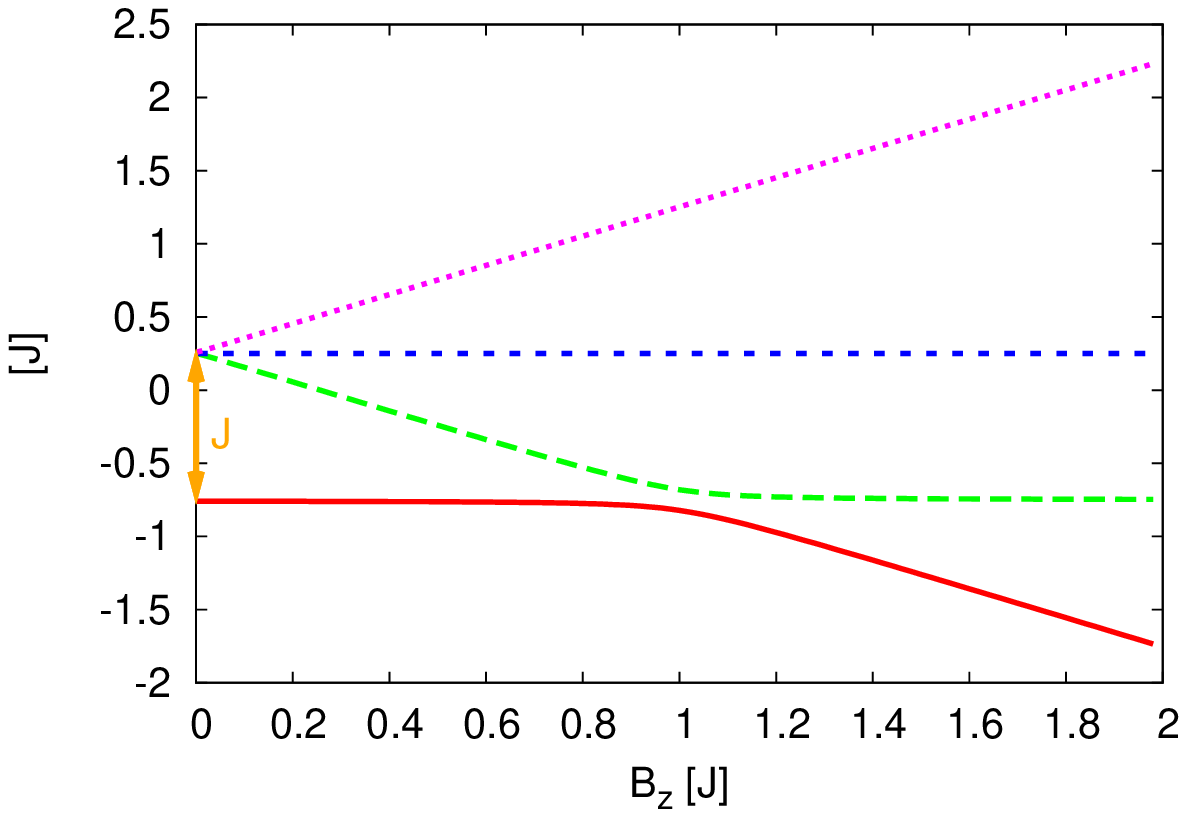}\label{fig:energies}}
  \subfigure[]{
    \includegraphics[height=0.31\linewidth]{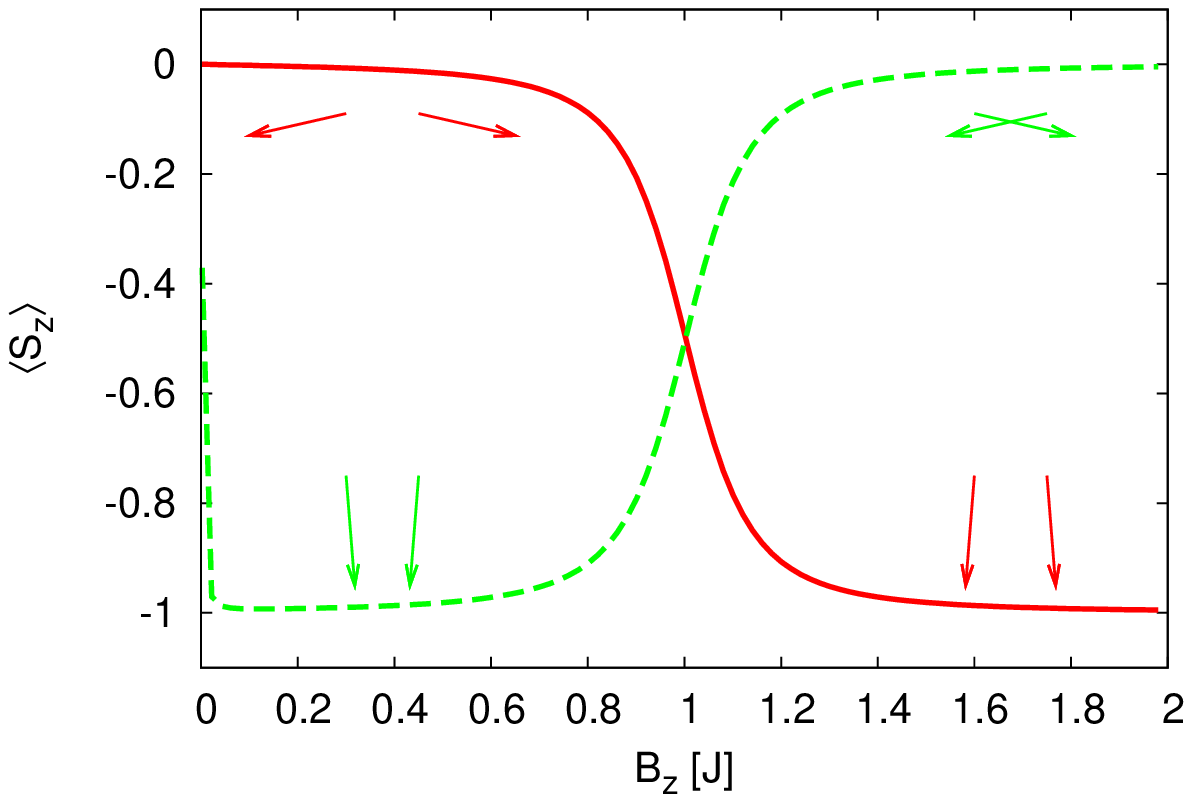}\label{fig:sz}}
  \caption{
    (a) Eigenenergies and (b) spin projection onto the magnetic field $\ve{B}$, which points along the z-axis and is perpendicular to the DM vector  $\ve{D}=(0.2\,J,0,0)$.
The red (green) curves and arrows in (a) and (b) refer to the ground (first excited) state.
The arrows in (b) indicate the relative orientation of the monomer spins.
The isotropic exchange $J$ is antiferromagnetic, resulting in near antiparallel orientation of the monomer spins in the ground state at low fields (red curves).
  \label{fig:eff}}
\end{figure}
The magnetic field $\mathbf{B}$ (in energy units, {{where $g\mu_B = 1$, assuming an isotropic $g=2$}}) enhances this mixing by bringing one triplet component close to the singlet. In general, this results in an energy level anticrossing at $|\ve{B}| \approx J$ (see \Fig{fig:energies}).
Interestingly, this mixing strongly depends on the orientation of the field $\ve{B}$ relative to the DM vector $\ve{D}$.
To see this, we quantize the spin along the magnetic field, i.e. $\ve{B}$ is chosen along the $z$-direction.
For $\ve{B}\,  \|\,  \ve{D}$ the singlet state does not mix with the lowest triplet and the states can cross,
since in this case the Zeeman and DM terms commute (they contain three different spin-components).
In contrast, for $\ve{B}  \perp\ve {D}$  the singlet and $S_z = \pm1$ triplet states are mixed.
As a result, when sweeping the magnetic field there is a smooth adiabatic change of the ground state magnetization instead of an instantaneous jump at the crossing point $|\ve{B}| \approx J$ (see \Fig{fig:sz}). Furthermore, the relative orientation of the two spins shows an interesting field dependence: For example, in case of the ground state the antiparallel spins are slightly canted 
(left of \Fig{fig:sz})
 at zero field and rotate towards each other with increasing field, until they are almost perpendicular at the anticrossing point $\ve{B} \approx J$. For higher fields $\ve{B} > J$ they become aligned by the external field
 (right of \Fig{fig:sz}).
The expectation value $\langle \mathbf{S_{}}_1 \times \mathbf{S}_2\rangle$ quantifies the ``canting'' of the spins. In particular, its sign equals the chirality characterizing the  intra-molecular spin structure.\par
In this work we predict the single electron transport signatures of the most elementary realization of DM interaction in a molecular dimer placed in a three terminal junction (see \Fig{fig:orbconf}). In section \ref{sec:model} we introduce a microscopic model which captures the basic physics discussed above and comment on the calculation of the transport current using a master equation. In section \ref{sec:normal} we present the transport current as function of both bias voltage and magnetic field. We show that the DM interaction leads to characteristic transport features which depend strongly on the gate voltage. We propose magnetic field measurements which allow for a determination of the magnitude of the $\ve{D}$-vector and the axis parallel to it. Finally in section \ref{sec:ferro} we present the non-linear conductance for polarized electrodes with antiparallel magnetizations. We show that due to the DM interaction the transport is sensitive to the reversal of both electrodes' polarizations. We explain how this additional measurement allows the polarity of the $\ve{D}$-vector to be determined as well.
\section{Model}\label{sec:model}
To predict the basic signatures of transport through a ``molecular double-dot'' as sketched in \Fig{fig:orbconf} including the effect of DM interaction a customary spin-Hamiltonian is insufficient: multiple charged states as well as orbital degrees of freedom of the monomers need to be accounted for.
Instead one has to consider the underlying many-body molecular Hamiltonian~\cite{Moriya60a} from which the \dm \ interaction (\ref{eq:eff}) derives:
$
  H_{\mathrm{m}}  =  H_{\mathrm{o}}  + H_{\mathrm{i}}   +  H_{\mathrm{so}} + H_{\mathrm{h}}+H_{\mathrm{Z}}$.
In particular, both ground ($i=1,2$) as well as excited orbitals of the monomers ($i=1',2'$) have to be included through
$ H_{\mathrm{o}} = \sum_{i, \sigma} \epsilon_{i} n_{i \sigma} $ with
energy $\epsilon_{i}$. The spin-orbit scattering with amplitude
$\lambda_i$ between these orbital states on monomer $i=1,2$ reads: $
H_{\mathrm{so}}= \sum_{i, j, \sigma, \sigma'} \lambda_{i} d_{i
\sigma}^{\dagger} \ve{l}_{ij} \cdot \mathbf{s}_{\sigma \sigma'} d_{j
\sigma'} $.  Here $ d_{i\sigma}$ and ${\hat n}_{i\sigma} =
d^{\dagger}_{i\sigma} d_{i\sigma}$ denote the electron and number
operator of orbital $i$ with spin $\sigma$ and $\mathbf{s}_{\sigma
\sigma'}$ are the spin $\half$ matrices.  We assume for simplicity no
orbital degeneracy implying that the orbital angular momentum is
quenched on each monomer ($\ve{l}_{ii}=0$). The off-diagonal matrix
elements of the orbital angular momentum $\ve{l}$, which parametrize the
local molecular orbital structure, are then imaginary and have a
crucial effect on the spin-orientation through the SO interaction.
The magnetic field tunes this effect through the Zeeman term $
H_{\mathrm{Z}} = \mathbf{B} \cdot \left( \frac{1}{2}\ve{l} + \ve{s}
\right) $, which acts on the spin $\ve s = \sum_{i,\sigma,\sigma'}
d^{\dagger}_{i\sigma} \mathbf{s}_{\sigma \sigma'} d_{i\sigma'}$ and
the orbital momentum $\ve l = \sum_{i,j,\sigma} d^{\dagger}_{i\sigma}
\mathbf{l}_{ij} d_{i\sigma}$.  Characteristic for a covalently bonded
dimer is the hybridization $ H_{\mathrm{h}} = \left(\sum_{i, j,
\sigma} d_{i \sigma}^{\dagger} t_{ij} d_{j \sigma} + h.c.\right) $
between the monomers and it couples their
excitations~\cite{Reckermann08b,Reckermann09a}.  The strong local
electron-electron interaction, $ H_\mathrm{\mathrm{i}} = \half
\sum_{\sigma, \sigma'} \sum_{ijkl} v_{i j k l}^{} d_{i
\sigma}^{\dagger} d_{j \sigma'}^{\dagger} d_{l \sigma'}^{} d_{k
\sigma}^{}$, plays a crucial role in several ways.  For the transport,
local repulsion results in Coulomb blockade of tunnelling and the
transport is dominated by single electron tunnelling (see below).
Furthermore, the direct (Hund's) exchange ($v_{ijji}=J_H$), combined
with antiferromagnetic superexchange induced by the hybridization
$H_{\mathrm{h}}$ and the local repulsion, gives rise to the isotropic
exchange interaction of the spins in \Eq{eq:eff} with $J = J_H +
\frac{t^2}{U}$ and $U=v_{iiii}-v_{ijij}$. The Hamiltonian $H_m$ is
diagonalized for different values of the electron charge $N$ of the
dimer ($N=0,1,2$).
\par
\begin{figure}[b!]  \subfigure[]{
    \includegraphics[height=0.27\linewidth]{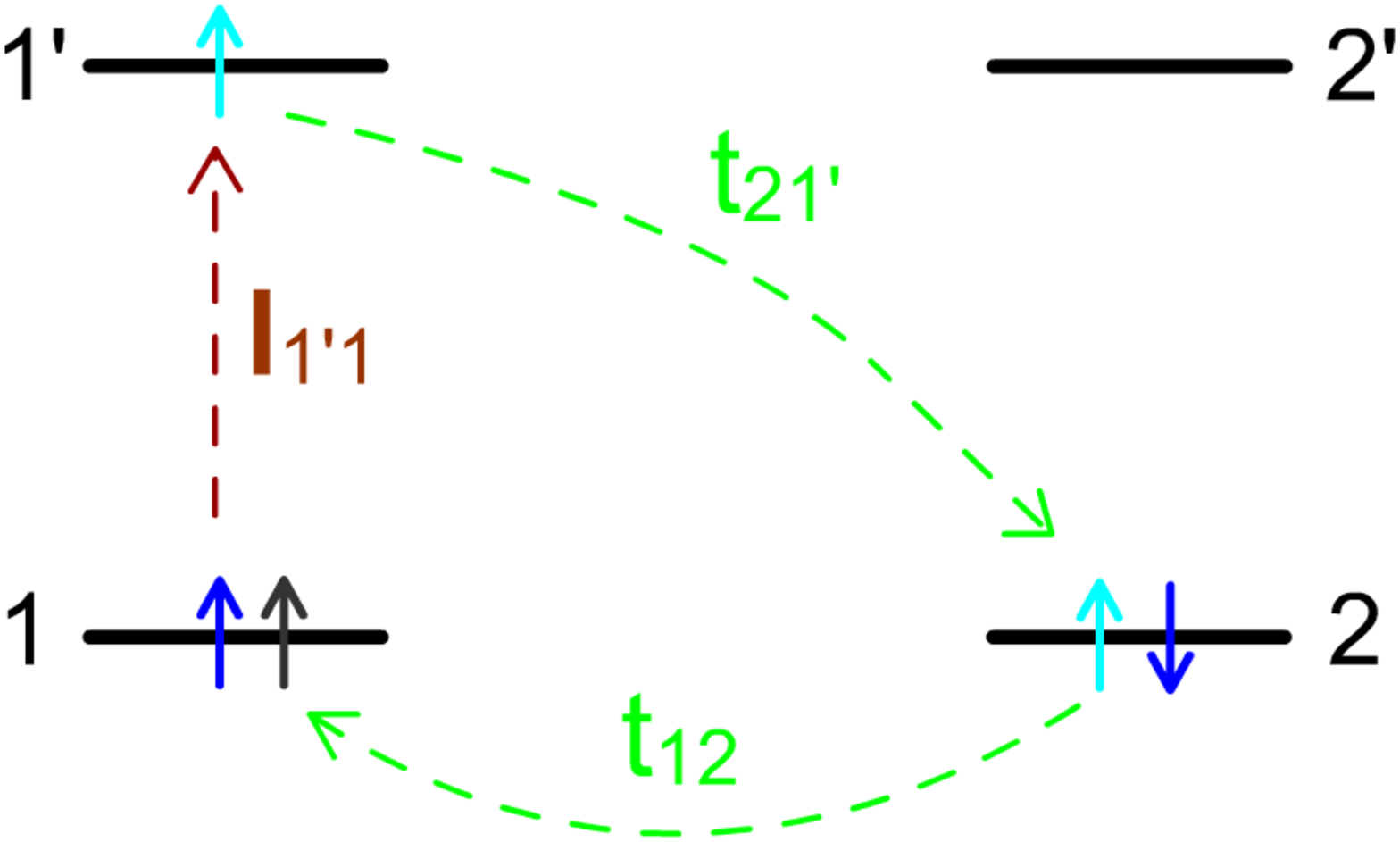}\label{fig:mag_sup}}
\subfigure[]{
    \includegraphics[height=0.27\linewidth]{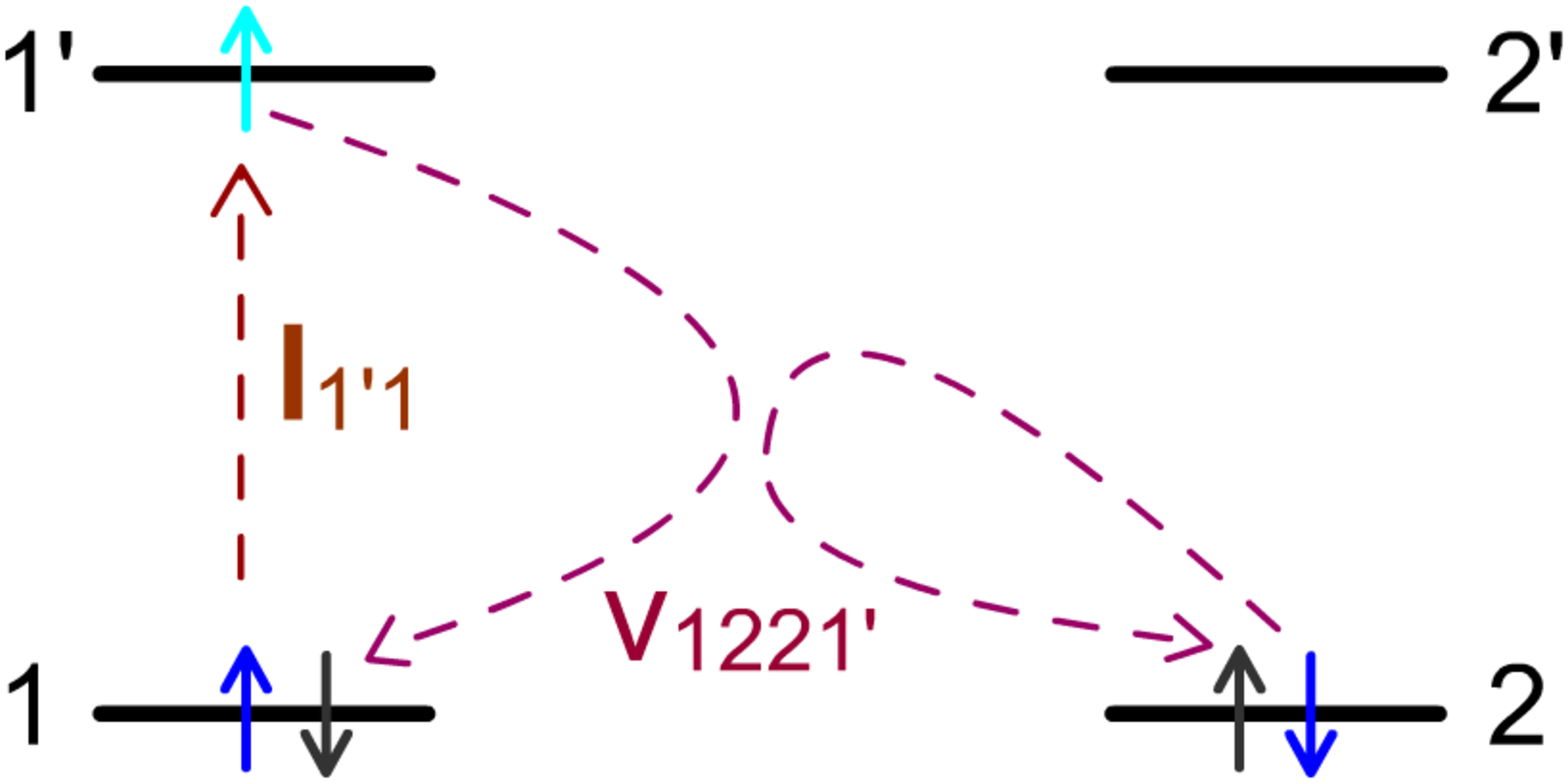}\label{fig:mag_dir}}
  \caption{Virtual processes leading to
Dzyaloshinskii-Moriya interaction. (a) superexchange and (b) direct
exchange.  }
\end{figure}

The above model incorporates the processes which generate the
Dzyaloshinskii-Moriya interaction in a way analogous to isotropic
exchange.  In linear order in spin-orbit coupling one can derive from
this the effective spin Hamiltonian \Eq{eq:eff} for $N=2$ with
\begin{eqnarray} \ve {D} &=& 2
  \frac{\mathrm{i}\lambda_1
\ve{l}_{1'1}}{\epsilon_{1'} - \epsilon_{1} } \left(\frac{t_{12}\,
t_{21'}}{U} +v_{1221'} \right)
   - (1 \leftrightarrow 2)
  \label{eq:D}
\end{eqnarray} where $1 \leftrightarrow 2$ denotes the same term with
the roles of the monomers interchanged.  Here virtual processes
involving tunnelling and spin-orbit scattering gives rise to the first
contribution (see \Fig{fig:mag_sup}). In addition, Coulomb excitation
of one monomer by the other through $v_{i j j k}$, sketched in
\Fig{fig:mag_dir}, induces an interaction of the same form,
contributing a second term.
Equation \Eq{eq:D} explicitly shows that a non-vanishing \dm \ vector is a joint effect of the asymmetries in the monomer excitation spectra, interaction matrix elements and spin-orbit couplings.
For definiteness we introduce the required asymmetry through the spin-orbit coupling on the monomers, $\lambda_{1} = 2 \lambda_{2}=$~8meV,
keeping other parameters symmetric.
Chemically, this asymmetry may be controlled by an appropriate choice of coordination ligands.
For molecular dimers a hierarchy of energy scales with $U,\epsilon_i > J,\lambda$ is expected.
The results discussed here, based on the parameters in \Tab{tab:par}, are representative of this order.
The corresponding energies in the effective spin-Hamiltonian are
\begin{equation}
J=4~\mathrm{meV}\qquad \ve{D}=-(0.32~\mathrm{meV},0,0)
\label{eq:param}
\end{equation}
i.e. the effective exchange $J$ is antiferromagnetic. By the choice of parameters processes indicated in \Fig{fig:mag_dir} do not contribute. We checked that they lead to the qualitatively similar results.
\par
\begin{table}[t!]
\caption{Model parameters. Double orbital occupation or occupation of an excited orbital is only virtually possible for these values. The resulting virtual spin scattering gives rise to the DM interaction.}\label{tab:par}
\begin{indented}
\item[]
\subtable[]{
\begin{tabular}{@{}lll}
\br
\multicolumn{3}{c}{\small One-particle interaction}\\
\mr
$\epsilon_{i}$ & 0 meV&$i \in (1,2)$\\
$\epsilon_{i'}$ & 50 meV&$i' \in (1',2')$\\
$t_{ij}$&10 meV&$j\neq i'$\\
$\ve{l}_{i'i}$&$(\mathrm{i},0,0)$&$i \in (1,2)$\\
\br
\end{tabular}}
\subtable[]{
\begin{tabular}{@{}lll}
\br
\multicolumn{3}{c}{\small Two-particle interaction}\\
\mr
$v_{iiii}$&125 meV&$i \in (1,1',2,2')$\\
$v_{ii'ii'}$&125 meV&$i \in (1,2)$\\
$v_{ijij}$&25 meV&$j \neq i'$\\
&&\\
\br
\end{tabular}
}
\end{indented}
\end{table}
Transport in three terminal molecular junctions~\cite{vanderZant06} can be modeled using a tunnel Hamiltonian approach,
parametrized by tunnel rate constants $\Gamma_{r}=\Gamma$ for tunnelling to electrode $r=L,R$.
For simplicity, these are assumed equal for the  ground and excited orbital.
Since the basic transport signatures of \dm \ have, to our knowledge, not been discussed before,
we focus on the basic single electron tunnelling transport regime (SET).
In our calculation we assume an electron temperature $T= 0.02~\mathrm{meV}$ in order to resolve the energy scale set by $\ve{D}$ in (\ref{eq:param}).
In the SET regime the stationary occupations of the molecular many-electron states and transport current can be calculated from a master equation \cite{Romeike06b}.
Some comments are in order.
First, off-diagonal elements of the density matrix in the molecular energy eigenbasis can in principle be important here,
since neither the spin length $S$ nor its projection $S_z$ are conserved quantities in tunnelling. These are not accounted for in the master equation. However, as we restrict our calculation of the tunnelling rates to the first order in the tunnel coupling and the energy spectrum contains no quasi-degeneracies, these elements can be neglected.
We have also checked this explicitly by comparing with results from a generalized master (or kinetic) equation accounting for non-diagonal elements.
\begin{figure}[b!]
\centering
    \includegraphics[height=0.6\linewidth]{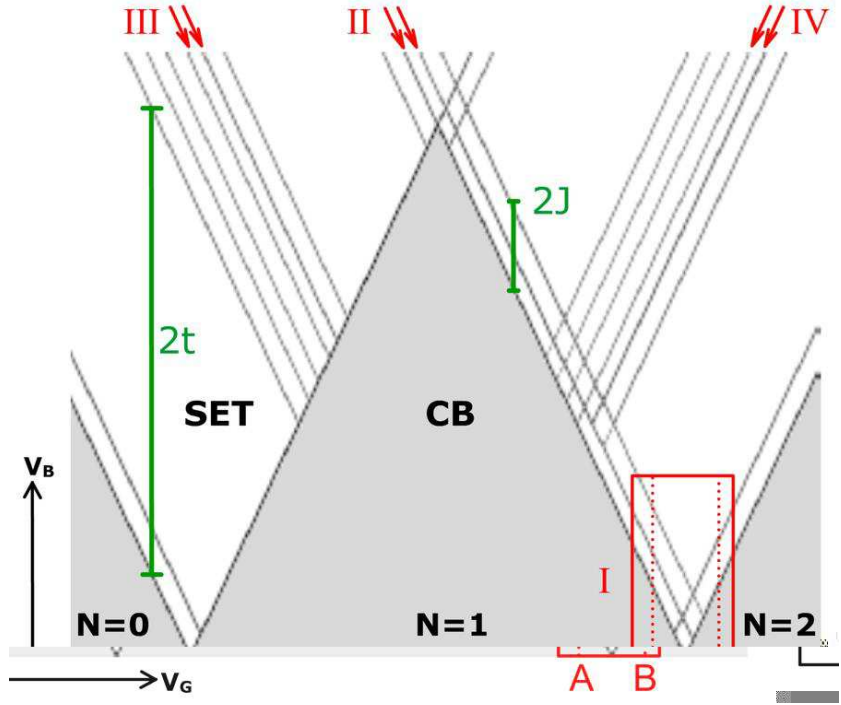}
  \caption{
    Schematic stability diagram as function of experimentally controllable voltages $V_B$ and $V_G$ for low magnetic field $B < J$ and no spin-orbit interaction. The grey-shaded area indicate the Coulomb blockade (CB) regime, where no current flows. Several SET resonances are marked by the box and the arrows. At these lines clear signatures of the \dm \ interaction develop when the spin-orbit interaction is included.} \label{fig:map}
\end{figure}
Secondly, we note that in magnetic molecules long spin-life times have been reported~\cite{Wernsdorfer00}.
For this reason we focus mainly on calculations neglecting any magnetic relaxation unrelated to transport
and find signatures of non-equilibrium magnetic states.
Still, to appreciate the possible influence of relaxation process (due e.g. coupling to junction phonons etc.), we have for each calculation performed a worst-case estimate calculation in which all excited states relax to all lower lying states within the same charge sector with a rate up to the tunnel rate (We assume a quadratic energy dependence of the density of states for these energy loss processes).
We thus identify transport signatures which are sensitive to relaxation and those which are robust against it.
\section{Transport signatures of DM interaction}
\subsection{Normal electrodes}\label{sec:normal}
For orientation, we show in \Fig{fig:map} the expected resonance lines for a dimer without DM interaction in a so-called stability diagram~\cite{Kouwenhoven97rev}, i.e. $dI/dV_B$ plotted versus the bias, $V_B$, and gate voltage, $V_G$.
We have explored this map in detail and identified significant transport signatures of Dzyaloshinskii-Moriya interaction along several sets of transitions, which are marked by I, II, III and IV in \Fig{fig:map}.
Due to the spin-orbit mixing  four additional electronic transitions can be observed which violate the spin selection rule $\Delta S_z = \pm \half$:
\begin{eqnarray}
  |\downarrow\rangle_\alpha \to |T_{+1}\rangle
  \,, \hspace{1cm}
 |\uparrow\rangle_\alpha   \to |T_{-1}\rangle
\end{eqnarray}
where the index $\alpha$ indicates whether the symmetric (regimes I and II) or the antisymmetric excited one-electron states (III and IV) are involved.
Since the latter transitions start from an excited state, they are sensitive to intra-molecular relaxation as our calculations bear out.
Apart from this the results in all regimes are qualitatively very similar.
\begin{figure}[t!]
\centering
  \includegraphics[height=0.4\linewidth]{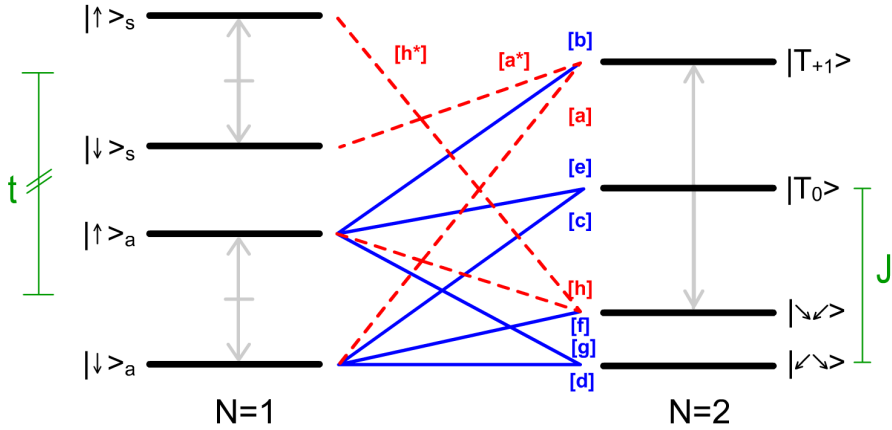}
  \caption{
    \label{fig:trans}
    Schematic many-body excitation spectra for $N=1$ and $N=2$ in a high magnetic field $B_z < J$
    relevant to regime I in \Fig{fig:map}.
    The energy scales of the hybridization $t$ and isotropic exchange $J$ are indicated. The grey arrows indicate the Zeeman splitting of the multiplets.
    The \dm \ interaction mixes the two low-lying $N=2$ states most strongly resulting in energy changes on the order of $|\ve{D}|\ll J,t$. The spin-orbit coupling also leads to spin-forbidden transitions (red dashed lines),
    in addition to the spin-allowed ones (blue solid lines). The shown transitions involving the symmetric excited one-electron states are labelled with an asterisk.
  }
\end{figure}
From hereon we focus on the regime I, where the situation is most clear since only charge states $N=1$ and $N=2$ of the dimer are involved.
Experimentally, this is the most relevant regime since the low voltage $V_B \sim J$ avoids complications due to e.g. junction instabilities.
In \Fig{fig:trans} we show the relevant transitions between the dimer states due to transport. The ground state is labelled with $\lch$ , i.e. indicating the intramolecular spin structure. It has mainly singlet character before the anticrossing and for high magnetic field closely resembles the pure polarized triplet state. For the first excited state $\rch$, which anticrosses with the ground state, it is the other way around (see \Fig{fig:sz}).
\par
\begin{figure}[t!]
  \subfigure[]{
    \includegraphics[height=0.5\linewidth]{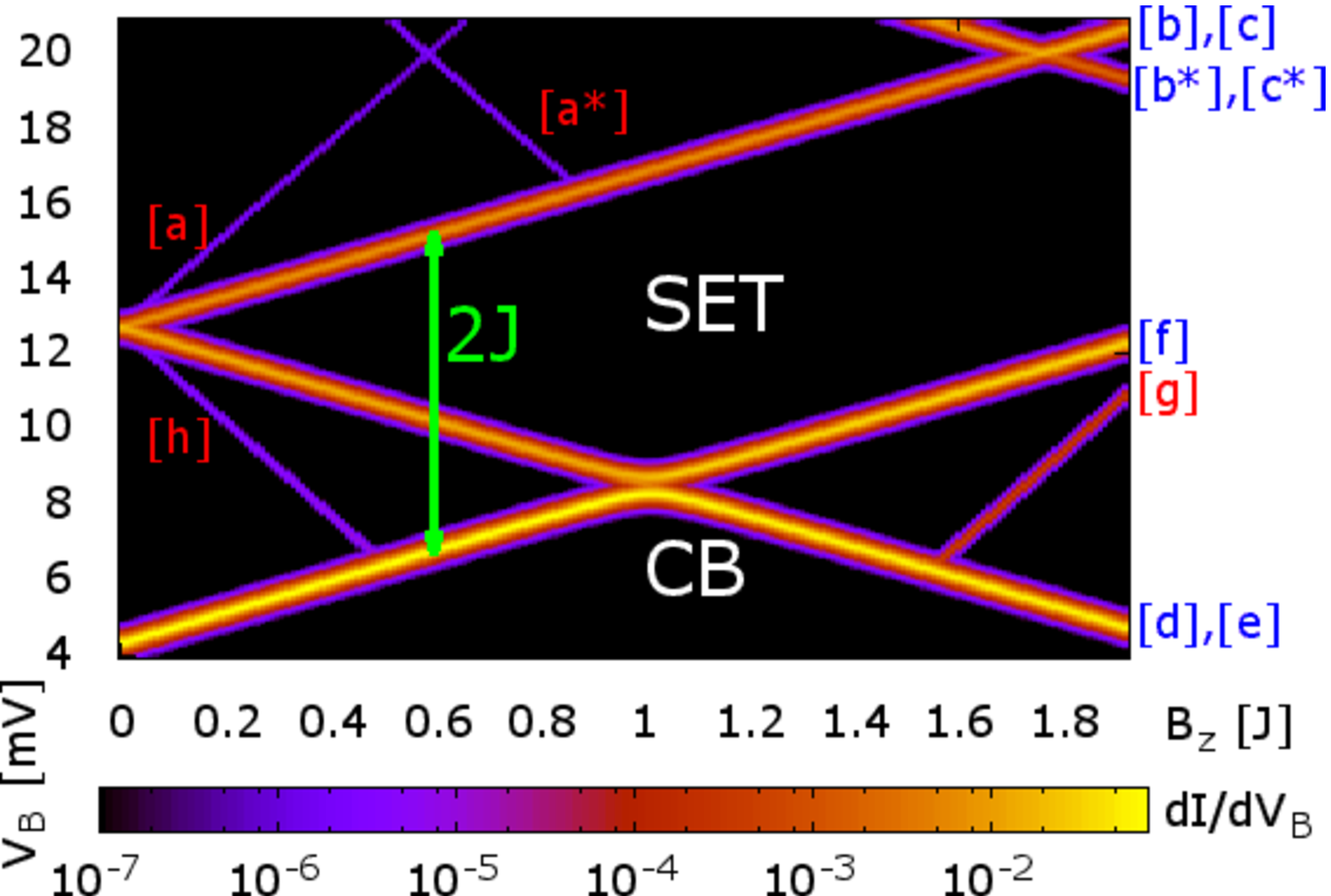}\label{fig:mfd_anti2}}
  \subfigure[]{
    \includegraphics[height=0.31\linewidth]{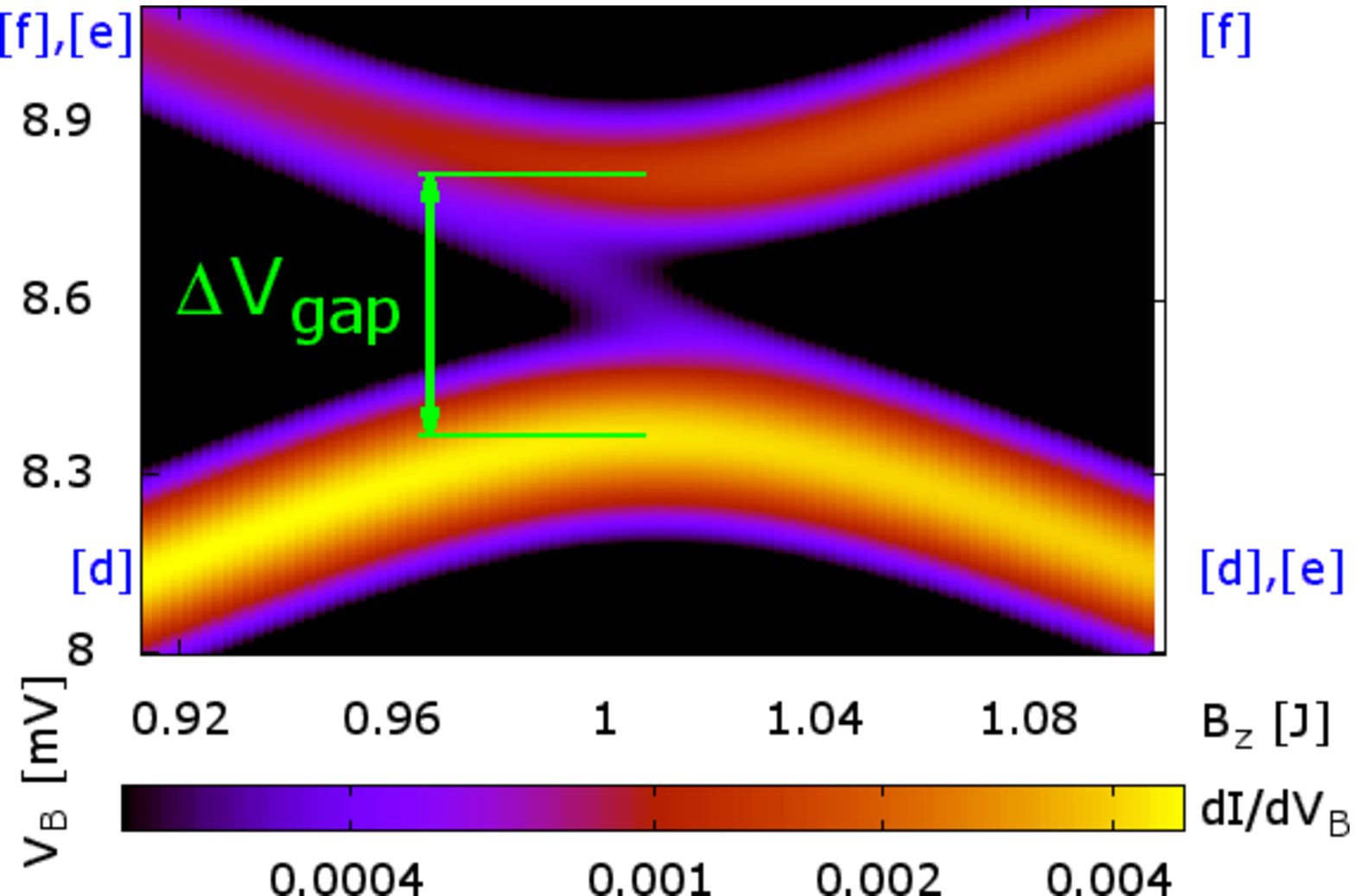}\label{fig:mfd_zoom}}
  \subfigure[]{
    \includegraphics[height=0.31\linewidth]{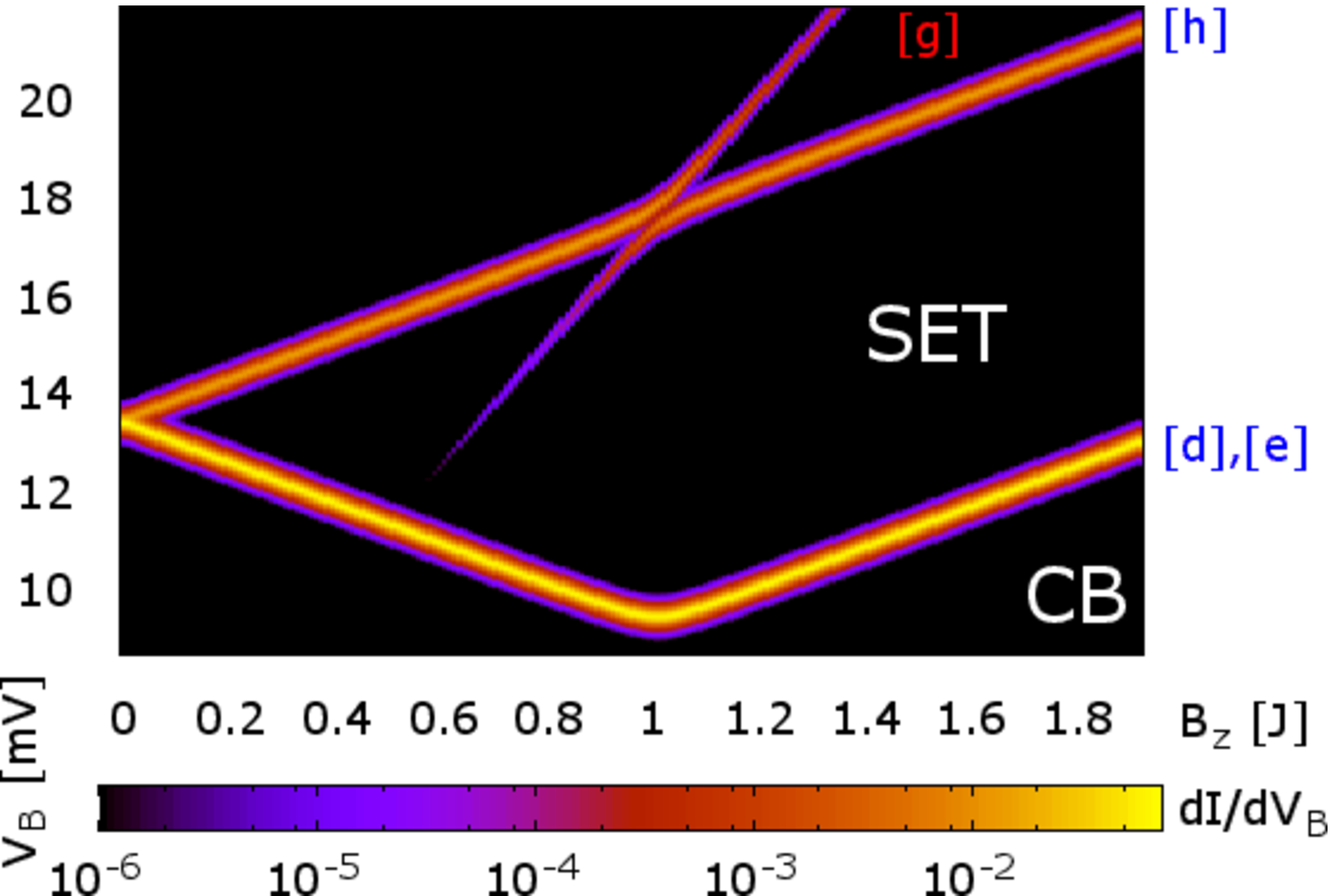}\label{fig:mfd_anti}}
  \caption{
    Magneto-conductance maps for regime I:
    dependence of $dI/dV_B$ and the magnetic field magnitude $B_z$ for $\ve{B} \perp \ve{D}$  .
    The conductance is normalized to $\Gamma$.
    (a) Overview of all transitions using a logarithmic scale  for a voltage sweep along line ``A'' in \Fig{fig:map}.
    The spin-forbidden transitions are weak and have an anomalous strong field dependence (triple $g$-factor). 
    The isotropic exchange constant $J$ can be extracted as indicated.
    (b) Zoom-in of the anticrossing gap in (a) revealing the line [e] running through the anticrossing.
    (c) Voltage sweep along line ``B'' in \Fig{fig:map}.
  }
\end{figure}

We first consider the case where $\ve{B} \perp \ve{D}$.
In \Fig{fig:mfd_anti2} we show $dI/dV_B$ as the bias $V_B$ (along line ``A'' in \Fig{fig:map}) and the magnetic field is swept.
Without spin-orbit coupling, one expects two degenerate transitions at low fields $|\ve{B}| \ll J$.
In contrast, two additional weak $dI/dV_B$ peaks (lines [a],[h]) are found with anomalous field dependence corresponding to a tripled $g$-factor since  $|\Delta S_z| =\frac{3}{2}$ is allowed by the spin-orbit interaction. This allows for their correct identification as due to breaking of a selection rule (instead of incorrectly assigning them to an additional state of the molecule). The intensity of these spin-forbidden transitions is proportional to $|\ve{D}|^4$.
\par
However, as the field is increased further the spin-allowed peak approaches the ground-state line (boundary of the SET regime) and the \emph{transport} spectrum displays a pronounced anticrossing.
This maps out the anticrossing in the $N=2$ \emph{energy} spectrum, as discussed in the introduction (cf. \Eq{eq:eff}).
The gap between the $dI/dV_B$ peaks directly gives access to the {\it magnitude} of the \dm \ vector:
\begin{eqnarray}
  \Delta V_{\mathrm{gap}} = \frac{1}{\sqrt{2}} |\ve{D}|\label{eq:Dgap}
\end{eqnarray}
However, the zoom in \Fig{fig:mfd_zoom} reveals an additional resonance running straight through the anticrossing gap.
This deviation from the energy spectrum is a true non-equilibrium transport effect.
This is confirmed by our calculations including relaxation rates:
this transition (marked by [e] in \Fig{fig:trans}) becomes suppressed since it starts in an excited state.
In contrast, the anticrossing  is robust against relaxation, since it involves transitions (marked by [d] and [f]) starting from the $N=1$ ground state.
\par
A similar signature of the \dm \ vector is found, if one measures $dI/dV_B$ as function of $V_B$ and $\ve{B}$ starting from the $N=2$ Coulomb blockade (CB) regime (line ``B'' in \Fig{fig:map}).
The calculated result is shown in \Fig{fig:mfd_anti}.
The boundary of the SET regime shows an adiabatic crossover, rather than a sharp transition, around $B_z \approx J$.
In addition an anticrossing of transport excitations appears with conductance amplitudes which show an unusual, pronounced bias and field dependence due to the singlet-triplet mixing.
For low field the excitation branch [g] is dominant since the transition is spin-allowed.
As one goes through the anticrossing this branch becomes suppressed since it has exchanged its physical character with branch [h] which is spin-forbidden at low fields.
We note that here the DM interaction gives rise to two effects: the anomalous g-factor of the spin-forbidden section of the branches and their anticrossing.
Finally, we mention that in the presence of relaxation the branch [h] becomes suppressed since the initial state is an excited state, in contrast to branch [g] which is unaffected.
\par
\begin{figure}[b!]
  \subfigure[]{
    \includegraphics[height=0.31\linewidth]{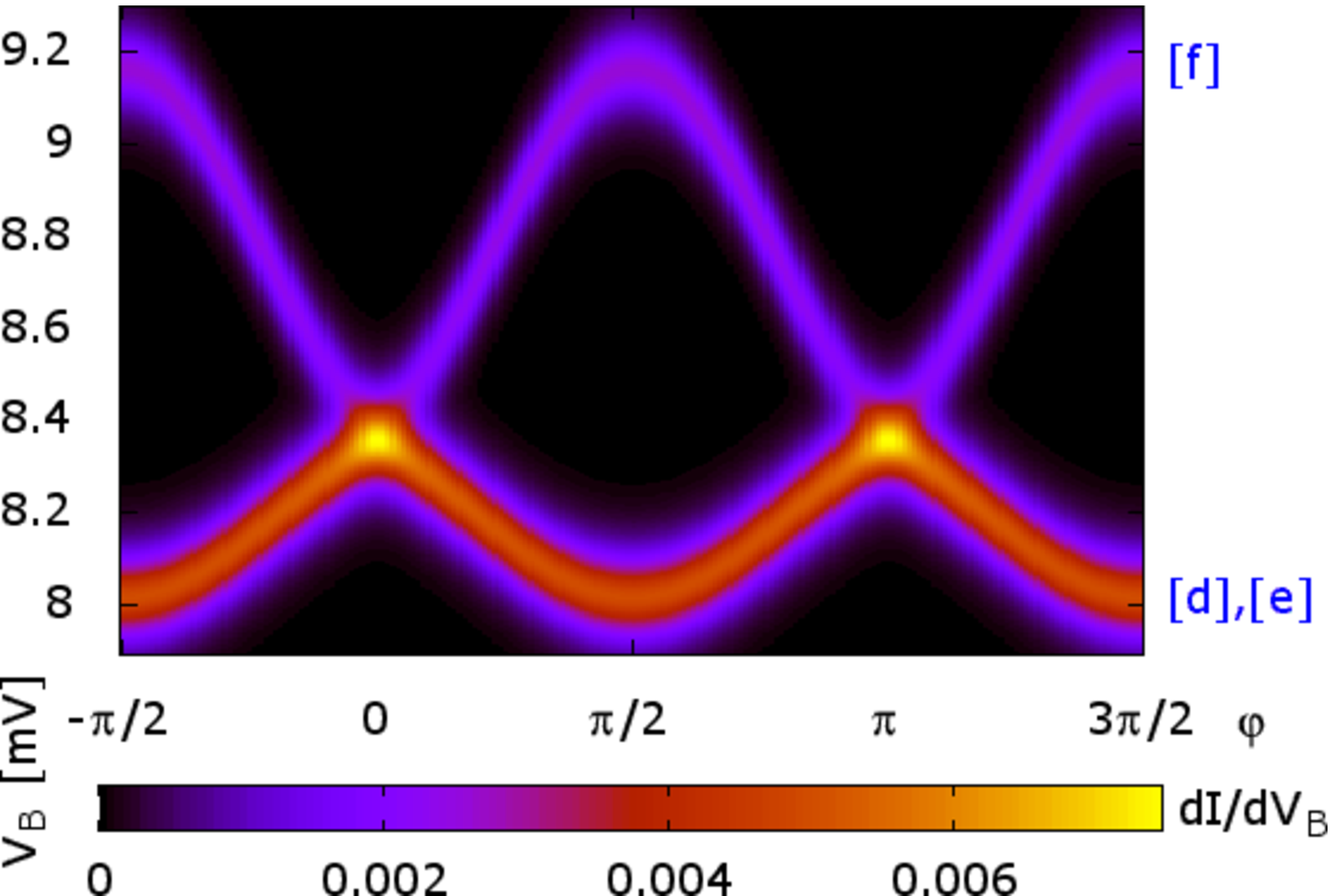}\label{fig:mfd_rotate}}
  \subfigure[]{
    \includegraphics[height=0.31\linewidth]{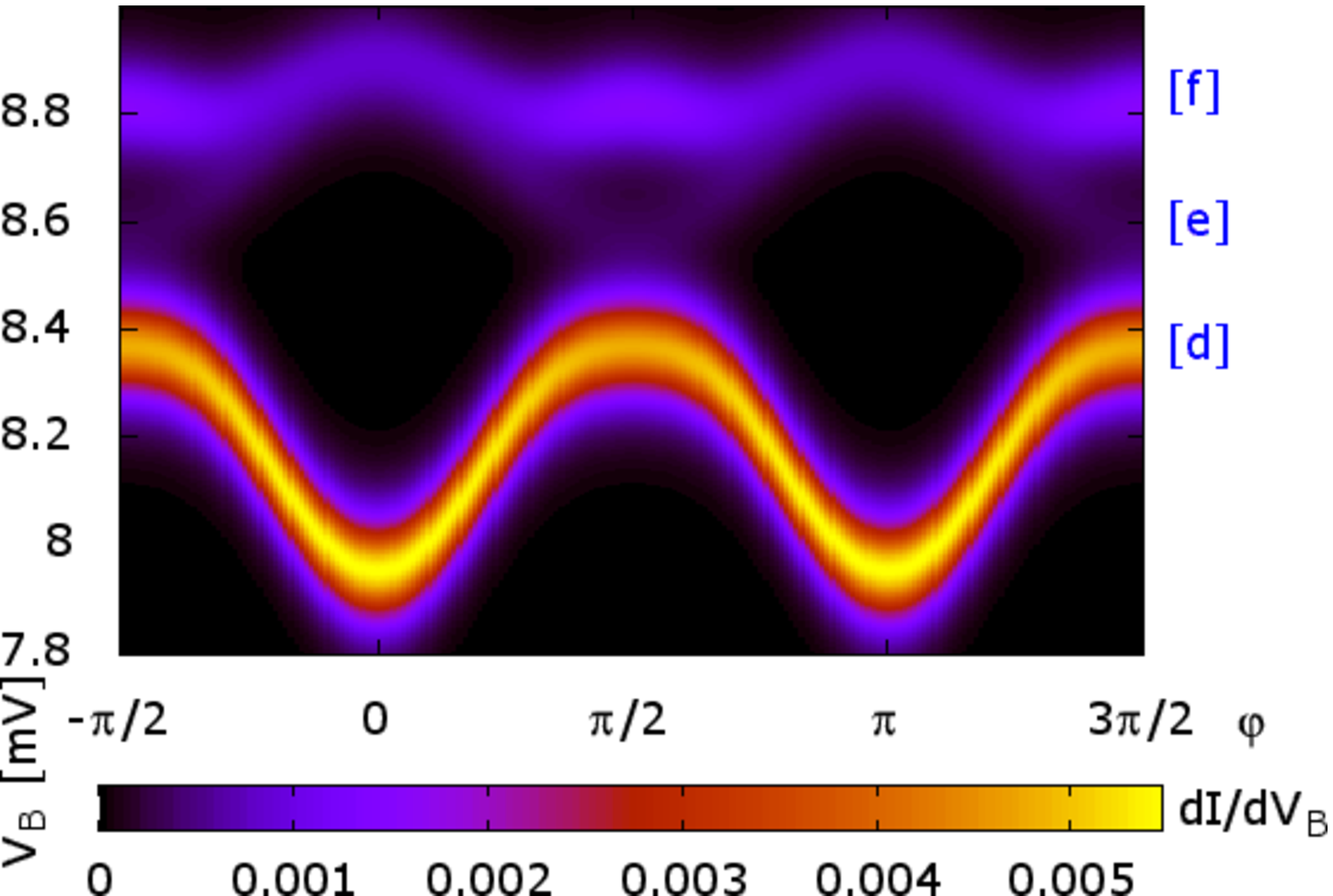}\label{fig:mfd_rotate2}}
  \caption{
    Magneto-conductance maps for regime I:
    dependence of $dI/dV_B$ on $V_B$ and the relative angle $\varphi$ between $\ve{B}$ and $\ve{D}$ for magnetic field magnitude (a) $B_z = 1.134 J$ and (b) $B_z = 1.08 J$.
  \label{fig:rotate}}
\end{figure}
We now discuss the effect of the magnetic field orientation
presented in figures~\ref{fig:mfd_rotate} and \ref{fig:mfd_rotate2}.
As explained in the introduction, for $\ve {B} \,\|\, \ve{D}$ the anticrossing in the energy spectrum vanishes.
Therefore if the external magnetic field is rotated, the size of the gap in the transport spectrum will oscillate
as \ref{fig:mfd_rotate} and \ref{fig:mfd_rotate2} show.
Interestingly, the non-equilibrium peak in the anticrossing gap (process [e]), shows also a clear dependence on the direction of the magnetic field due to the admixture between $\ket{S}$ and $\ket{T_0}$ for $\ve {B} \,\|\, \ve{D}$.
However, when rotating the magnetic field, the field magnitude for which singlet and triplet should (anti-)cross is shifted. This is due to an interplay between \emph{orbital} Zeeman effect and spin-orbit coupling leading to an anisotropic renormalization of the effective $g$-tensor in \Eq{eq:eff}. This effect is captured by our exact diagonalization of the model $H_m$.
Furthermore, the {orbital} Zeeman effect induces oscillations in the energy and transport spectrum when rotating the field.
Therefore the observed oscillations with period $\pi$ in figures~\ref{fig:mfd_rotate},\ref{fig:mfd_rotate2} cannot simply be attributed to the DM interaction alone.
This underlines the importance of transport calculations using the detailed model $H_m$.
\par
We now explain how the above transport signatures allow one to determine the \emph{axis} of the DM vector experimentally.
Starting with an arbitrary orientation of the magnetic field, one first measures the dependence on the field magnitude  in search for an anticrossing resulting in a map similar to \Fig{fig:mfd_anti2}.
Having found the anticrossing, one then fixes the field strength and rotates the field by $2\pi$.
One then observes an oscillation of the anisotropic magneto-conductance similar to that shown in \Fig{fig:mfd_rotate2} with a remnant gap at the crossing points.
Next one can reduce this remnant gap to zero by a combination of adjusting the magnetic field (to compensate the g-factor renormalization) and changing the plane of rotation until it includes the $\ve{D}$-vector.
One has determined the axis of the $\ve{D}$ vector when the measured map corresponds to \Fig{fig:mfd_rotate} where for $\ve{B} \,\|\, \ve{D}$ a crossing is observed.
We emphasize that the maximal gap observed in this figure for $\varphi=\pi/2$ does \emph{not} allow the DM vector magnitude to be read off.
Instead, one needs to orient the magnetic field perpendicular to the $\ve{D}$-vector
and measure the map corresponding to \Fig{fig:mfd_anti2}.
The measured anticrossing gap  directly gives the magnitude of the $\ve{D}$-vector via \Eq{eq:Dgap}.
We note that this last step is necessary due the renormalization of the $g$-factor.
\subsection{Ferromagnetic electrodes}\label{sec:ferro}

Although with the measuring scheme proposed above one can determine the axis of the $\ve{D}$-vector, it is impossible to infer whether it is aligned antiparallel or parallel to this axis. This is due to the fact that
the effective Hamiltonian (\ref{eq:eff}) is invariant under simultaneous reversal of $\ve{D}$ and interchange of the spins $\ve{S}_1$ and $\ve{S}_2$ (leaving the electrodes unaltered).
Clearly, with ferromagnetic leads this symmetry is broken and one can therefore probe the chirality of the ground state.
Although nanoscale spin-valve structures can be realized nowadays~\cite{Pasupathy04kondo,Bolotin06,Hauptmann08},
we note that to perform measurements with polarized electrodes in a magnetic field in the tesla range one may make use of alloys with large coercive fields and the effect of micromagnetic structure of the tunnel contact region~\cite{Bernard-Mantel09}.

\begin{figure}[t!]
\centering
  \subfigure[]{
    \includegraphics[height=0.13\linewidth]{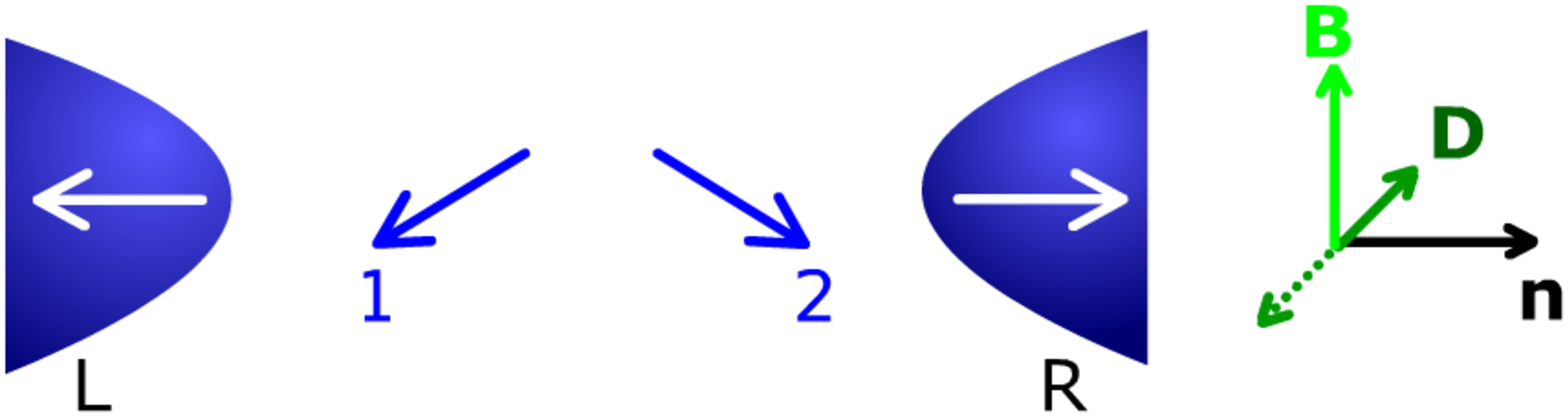}\label{fig:ferro_favour}}
  \subfigure[]{
    \includegraphics[height=0.13\linewidth]{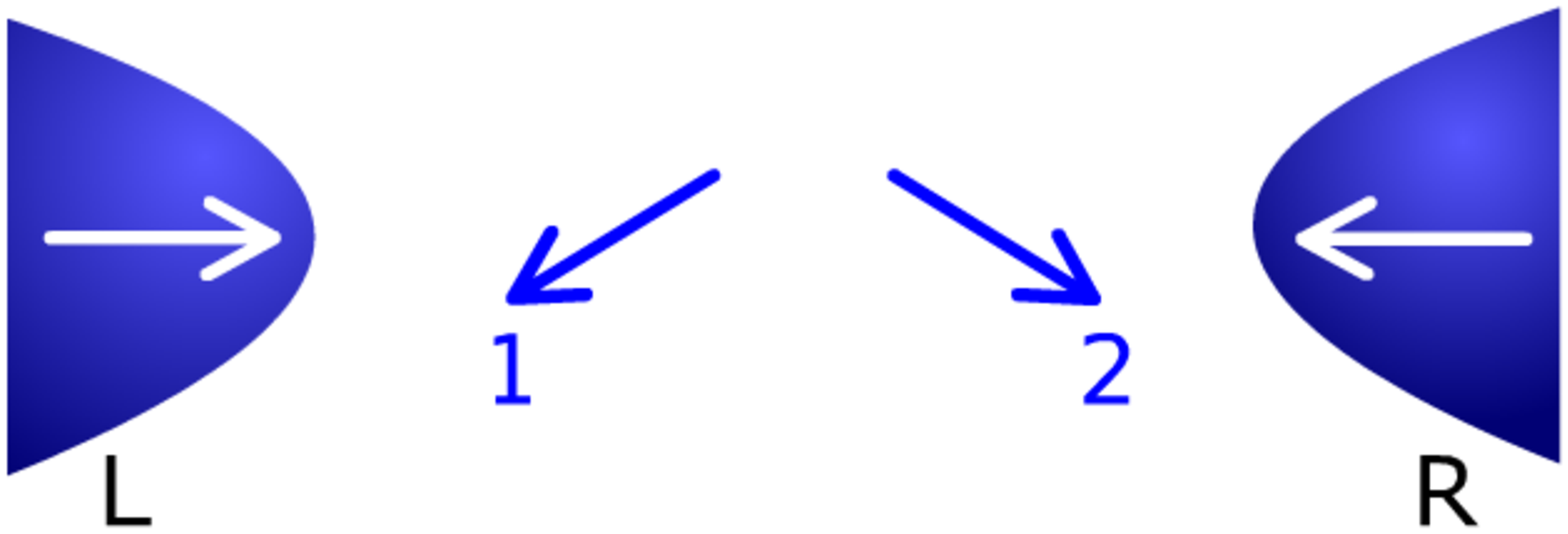}\label{fig:ferro_unfavour}}
  \caption{
    Probing the chirality of the ground state with ferromagnetic leads with polarization $p$ (right) and $-p$ (left). (a) $p>0$: for each monomer the projection of its spin onto the polarization vector of the adjacent electrode is positive and the  current is not inhibited. (b) $p<0$: reversal of the polarizations suppresses the current relative to situation in (a).
}
\end{figure}

To clearly probe the chirality both polarization axes
of the electrodes needs to be perpendicular to the plane spanned by the external magnetic field and the $\ve{D}$-vector. To obtain a maximal effect the field needs to be perpendicular to the axis of the DM vector, i.e., $\ve{B} \perp \ve{D}$.
Note that these requirements can be met in principle since the directions can be determined by first applying the measurement scheme of the previous section to the case ferromagnetic electrodes.
We make use of the fact that the tunnelling onto and off the monomers is sensitive to the relative orientation of the monomer spin to the polarization vector of the connected electrode.
Therefore, for antiparallel lead polarizations the monomer spins in the $N=2$ ground state will have orientations which are either both favourable (\Fig{fig:ferro_favour}) or both unfavourable for the transport (\Fig{fig:ferro_unfavour}).
The transport measured for two opposite antiparallel polarizations is thus expected to probe the chirality of the spin-canting due to the DM interaction.

\begin{figure}[b!]
    \includegraphics[height=0.45\linewidth]{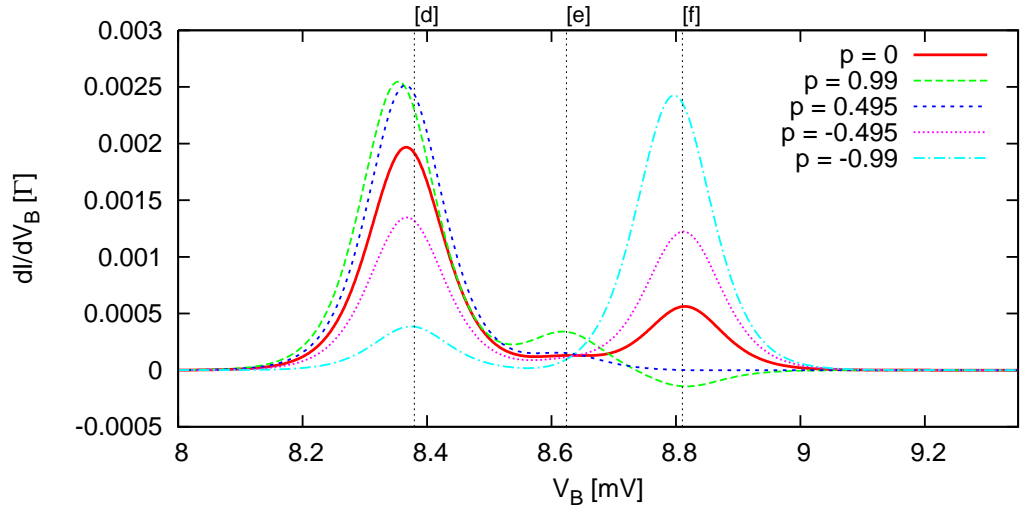}
  \caption{
  Conductance $dI/dV_{B}$ as function of $V_{B}$ along line ``A'' in \Fig{fig:map} for fixed magnetic field $\ve{B} \perp \ve{D}$ and $B_z \approx J$. The green vertical lines indicate the transition energies.
The small shifts of the [d] and [f] peaks is due to the combined effect of the large difference in the tunnel rates for these states and the finite temperature (see \cite{Bonet2002a,Romeike06b}).
Note that the non-equilibrium mid-gap peak [e] discussed in section \ref{sec:normal} is sensitive to the polarization as well.
}\label{fig:ferro_calc}
\end{figure}

In \Fig{fig:ferro_calc} we present $dI/dV_B$ vs $V_B$ curves calculated along line ''A'' in \Fig{fig:map} at the anticrossing ($|\ve{B}| = J$) for different degrees and sign of the polarizations while maintaining the antiparallel configuration.
The degree of polarization for the right electrode is defined as
$p = (\nu_{\uparrow}-\nu_{\downarrow})/(\nu_{\uparrow}+\nu_{\downarrow})$
where $\sigma=\uparrow,\downarrow$ denotes the spin quantized in the direction of the polarization vector $\ve{n}$ of the right lead
and $\nu_\sigma$ denotes the density of states for spin $\sigma$.
For the antiparallel configuration considered here the degree of polarization of the left electrode equals $-p$.
Figure~\ref{fig:ferro_calc} shows that the ground to ground conductance peak [d] is clearly enhanced for $p>0$ and suppressed for $p<0$, whereas the peak [f] involving the $N=2$ excited state with opposite chirality shows the opposite behaviour.
This effect is a clear signature of the definite spin-chirality induced by the DM interaction.
The effect is most pronounced near the anticrossing where the spin-multiplets strongly mix due to DM interaction,
and does not even require nearly complete polarization of the electrodes ($|p| \approx 1$).
We mention that even at fields below the anticrossing the effect can be significant if the polarization is sufficient.

From a measurement resembling \Fig{fig:ferro_calc} one can experimentally conclude that the ground state spin structure is as shown in \Fig{fig:ferro_favour}. Given the polarization $\ve{n}$ of the right electrode, one infers that the $\ve{D}$ vector points in the direction $\pm \ve{B} \times \ve{n}$ if an enhancement/suppression of the ground to ground state peak is measured.
From this one then finds the microscopic spin-canting $\ve{S}_1 \times \ve{S}_2$ in the $N=2$ ground state, which is antiparallel to $\ve{D}$ in order to minimize the DM interaction energy.

\section{Conclusion}
We have predicted non-linear transport signatures of the \dm \ interaction for the generic case of a molecular dimer.
We have shown how both the magnitude and the orientation of the vector characterizing this interaction can be determined from three-terminal transport measurements. 
The violation of the spin selection rules results in weak spin-forbidden electronic transitions with an apparent tripled $g$-factor in the transport spectrum. Most notably, the adiabatic change in the magnetization induced by the DM interaction causes a strong dependence of conductance peak amplitudes on the field strength and orientation. We show that when extracting the axis of the DM-vector from the magneto-conductance map, one has to account for the orbital Zeeman effect and the resulting anisotropic renormalization of the $g$-factor.
The proposed measurement scheme is robust against significant relaxation of the magnetic excitations.
When probing the dimer with polarized electrodes with antiparallel magnetizations, the canted spin structure induced by the DM interaction leads to a strong sensitivity to the reversal of the polarization vectors.
This effect allows the \emph{absolute orientation} of the DM vector to be inferred.
\par
On a more general level, this work shows how transport can probe complex intramolecular excitations
which arise when monomers hybridize and their local excitations become coupled.
In analogy to this work, mechanical excitations of two coupled monomers were shown~\cite{Reckermann08b} to result in SET transport effects signalling pseudo Jahn-Teller dynamics and the breakdown of the Born-Oppenheimer approximation.
For the three-electron dimer states an additional interplay of these vibrations with the monomer spins resulting from  double-exchange allows for vibrational spin-spectroscopy and even results in a vibration-induced spin-blockade~\cite{Reckermann09a}.
All this goes to show that the quantized degrees of freedom of molecular ``double quantum dots'' and their various interactions lead to a broad range of non-linear transport effects which provide detailed insight and may prove useful in nanoelectronic devices.
\par
We thank M. Leijnse, F. Reckermann for technical support
and S. Andergassen, S. Das, A. Bernard-Mantel, N. Konstantinidis, P. Dederichs, S. Lounis,  P. K\"ogerler for stimulating discussions.
\bibliographystyle{unsrt}
\bibliography{paper}
\end{document}